\documentclass[twocolumn]{aastex631}

\usepackage{lineno}
\usepackage{multirow}

\received{March 1, 2023}
\revised{April 1, 2023}
\accepted{\today}

\begin{document}

\title{Age distribution of exoplanet host stars: Chemical and Kinematics age proxies from GAIA DR3}

\correspondingauthor{C. Swastik}
\email{swastik.chowbay@iiap.res.in}

\author[0000-0003-1371-8890]{C. Swastik}
\affiliation{Indian Institute of Astrophysics, Koramangala 2nd Block, Bangalore 560034, India}
\affiliation{Pondicherry University, R.V. Nagar, Kalapet, 605014, Puducherry, India}

\author[0000-0003-0799-969X]{Ravinder K. Banyal}
\affiliation{Indian Institute of Astrophysics, Koramangala 2nd Block, Bangalore 560034, India}

\author[0000-0002-0554-1151]{Mayank Narang}
\affiliation{Academia Sinica Institute of Astronomy $\&$ Astrophysics, 11F of Astro-Math Bldg., No. 1, Sec. 4, Roosevelt Road, Taipei 10617, Taiwan, Republic of China}

\author[0000-0001-6093-5455]{Athira Unni}
\affiliation{Indian Institute of Astrophysics, Koramangala 2nd Block, Bangalore 560034, India}
\affiliation{Pondicherry University, R.V. Nagar, Kalapet, 605014, Puducherry, India}

\author[0000-0001-8075-3819]{Bihan Banerjee}
\affiliation{Department of Astronomy and Astrophysics, Tata Institute of Fundamental Research
Homi Bhabha Road, Colaba, Mumbai 400005, India}

\author[0000-0002-3530-304X]{P. Manoj}
\affiliation{Department of Astronomy and Astrophysics, Tata Institute of Fundamental Research
Homi Bhabha Road, Colaba, Mumbai 400005, India}

\author[0000-0003-0891-8994]{T. Sivarani}
\affiliation{Indian Institute of Astrophysics, Koramangala 2nd Block, Bangalore 560034, India}

\begin{abstract}
The GAIA space mission is impacting astronomy in many significant ways by providing a uniform, homogeneous and precise data set for over 1 billion stars and other celestial objects in the Milky Way and beyond. Exoplanet science has greatly benefited from the unprecedented accuracy of stellar parameters obtained from GAIA.  In this study, we combine photometric, astrometric, and spectroscopic data from the most recent Gaia DR3 to examine the kinematic and chemical age proxies for a large sample of 2611 exoplanets hosting stars whose parameters have been determined uniformly. Using spectroscopic data from the Radial Velocity Spectrometer (RVS) onboard GAIA, we show that stars hosting massive planets are metal-rich and $\alpha$-poor in comparison to stars hosting small planets. The kinematic analysis of the sample reveals that the stellar systems with small planets and those with giant planets differ in key aspects of galactic space velocity and orbital parameters, which are indicative of age. We find that the galactic orbital parameters have a statistically significant difference of 0.06 kpc for $Z_{max}$ and 0.03 for eccentricity respectively. Furthermore, we estimated the stellar ages of the sample using the MIST-MESA isochrone models. The ages and its proxies for the planet-hosting stars indicate that the hosts of giant planetary systems are younger compared to the population of stars harboring small planets. These age trends are also consistent with the chemical evolution of the galaxy and the formation of giant planets from the core-accretion process.
\end{abstract}

\section{Introduction}
Unprecedented advancements in astronomy and astrophysics are being achieved through the GAIA space mission. This is made possible by the sheer volume and quality of data obtained from high-precision spectroscopic, astrometric, and photometric instruments on board GAIA  \citep{GAIA,2022arXiv220800211G}. GAIA's impact on exoplanet science is no less impressive. The planet  discovery potential of the GAIA  mission was first investigated by \cite{2014ApJ...797...14P}. Astrometry being an important detection technique, can provide  both the mass and orbital period of the planet.  In the next data release, GAIA is expected to detect several thousand  new exoplanets, thanks to its 30-fold increase in astrometric precision compared to its predecessor, HIPPARCOS \citep{1997SSRv...81..201V}. That said, GAIA's contribution to exoplanet science goes beyond detecting planets by astrometry. 

The GAIA has enabled deriving stellar and planetary radii with the highest possible accuracy ($\sim$5$\%$) using the most precise parallaxes to date \citep{2018ApJ...866...99B,2020AJ....160..108B}. Accurate stellar and planetary radii have thus helped to solidify star-planet correlations such as planet-radius versus stellar metallicity \citep{2014Natur.509..593B,2018ApJ...853...37S,nar18}. Further, the existence of the ``radius valley'' a gap in the distribution of exoplanet radii that separates the super-earths ($R \sim 1.4 R_{\oplus}$) and mini neptunes ($R \sim 2.4 R_{\oplus}$), with a clear paucity around $R \sim 1.8 R_{\oplus}$, is now well established from observational results \citep{2017AJ....154..109F,2019ApJ...880L...1A,2022AJ....163..179P}. 
%Thus, GAIA's mission have contributed greatly in understanding the demographics of exoplanets. 

Significant research effort is also devoted to exploring star-planet connections, showing how the fundamental properties of stars determine the orbital and physical characteristics of the planetary systems.  For example, spectroscopic studies have shown that the metallicity distribution for stars with small ($M_{P}<$0.3$M_{J}$) and giant planets (0.3$M_{J}\leq M_{P}\leq 13M_{J}$) is different, indicating that they likely belong to different populations and also metallicity plays a key role in giant planet formation \citep{1997MNRAS.285..403G, san01, fis05, 2007ARA&A..45..397U, 2012Natur.486..375B,2014Natur.509..593B,2014ApJ...789L...3D,2015AJ....149..143F,2017AJ....154..108J, pet18, mul18,nar18, 2021AJ....161..114S}. Further, a detailed abundance analysis shows that the chemical composition of stars hosting small and giant planets is different, with the latter being $\alpha$-poor \citep{2022AJ....164...60S,2022arXiv220810057U}. Using   [$\alpha$/Fe] ratio as a proxy for the age, these studies suggest that the small planetary systems may have started forming early in the Milky Way history compared to the late formation onset of the giant planets \citep{2019A&A...624A..78D}. Estimating the ages from isochrone fittings for a sub-sample of exoplanet hosting stars have also arrived at similar conclusions \citep{2015A&A...575A..18B,2022AJ....164...60S}. Such age differences are also reported based on the position and kinematics studies of the confirmed population of planet-hosting stars (Narang et al under review).

\begin{figure}[t]
\centering
\includegraphics[width=1\columnwidth]{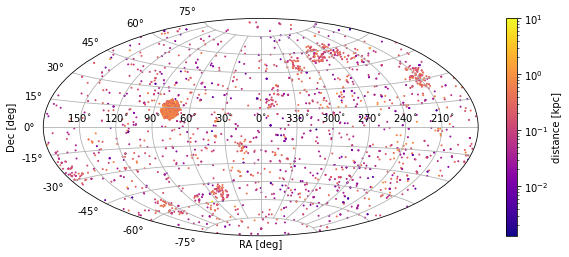}
\caption{Exoplanet hosting stars for which Gaia parameters are available. The colorbar represents the distance in kpc from the sun. Also, note the blob of the planet above the galactic plane which represents the Kepler field.}
\label{f1}
\end{figure}

\begin{figure}[t]
\centering
\includegraphics[width=1\columnwidth]{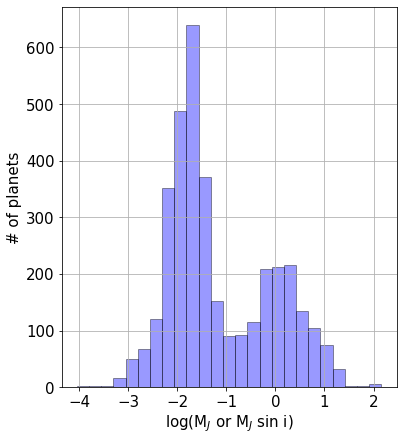}
\caption{Planet Mass Distribution for the Sample with Host Stars Listed in GAIA Archive.}
\label{plmass}
\end{figure}

\begin{figure}[t]
\centering
\includegraphics[width=1\columnwidth]{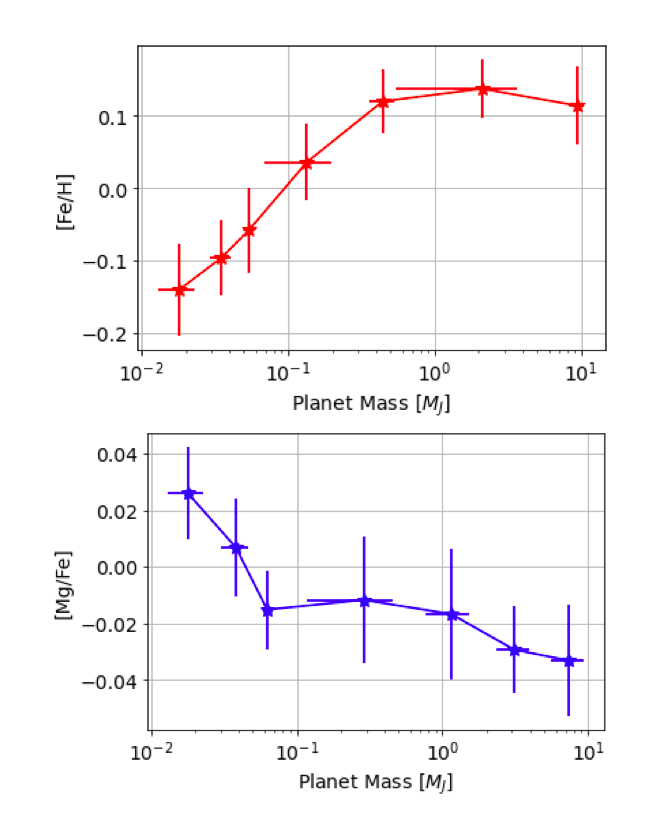}
\caption{\textbf{Top :} Host star metallicity as a function of planet mass.  \textbf{Bottom :} [Mg/Fe] of planet-hosting stars as a function of planet mass. The errors in metallicity and Mg abundances are represented by the standard error of the mean, whereas the errors in planet mass are represented by the standard deviation in each bin.}
\label{f2}
\end{figure} 

Important as they are, the majority of these studies used  mixed samples of stars originally  observed in different planet search and follow-up surveys \citep{2011A&A...526A.112S,2016ApJS..225...32B,2018ApJS..237...38B}. In most cases, the stellar properties themselves are determined using  different observing strategies, instrument settings and analysis methods.  This results in various systemics and offsets, making the interpretation and comparison more difficult across different studies. Ideally, to make the findings more robust and universal,  a sufficiently large sample of stars should be observed with the same equipment under similar conditions, and a uniform methodology must be applied to determine the parameters of interest. With the latest release of the GAIA DR3 data, it is possible to study a much larger and uniform sample of planet-hosting stars whose properties have been determined homogeneously. 

The General Stellar Parametriser (GSP) module uses spectra from a medium resolution (R $\sim$~11,500) radial velocity spectrograph \citep{2022arXiv220605541R}. The GSP-Spec module computes the stellar atmospheric parameters ($T_{eff}$, log$\:$g, metallicity ([M/H])) and abundances ([X/Fe]) for thirteen different species for each star, including three Fe-peak elements, Cr, Ni and Fe. Additionally, it provides the mean abundances of eight $\alpha$-elements (O, Ne, Mg, Si, S, Ar, Ca, and Ti) in the catalogue. Other than spectroscopic parameters, GAIA also provides homogeneous and accurate astrometric and photometric parameters for nearly two billion stars, the largest to date.

In this paper, we investigate a sample of 2611 planet-hosting stars whose parameters have been determined homogeneously. We used the Gaia DR3 data and analyzed the spectroscopic and kinematic parameters of stars hosting small and giant planets. The outline of this paper is as follows:  We describe our sample in Section~\ref{s2}. We report the results of the spectroscopic, kinematic, and isochrone age analyses in Section~\ref{s3}. In Section~\ref{s4}, we briefly discuss our results in the context of planet formation theories and also elude to possible  biases and systematic affecting our findings. Finally, in Section~\ref{s5}, we conclude and summarise the results.

\begin{figure}[t]
\centering
\includegraphics[width=1\columnwidth]{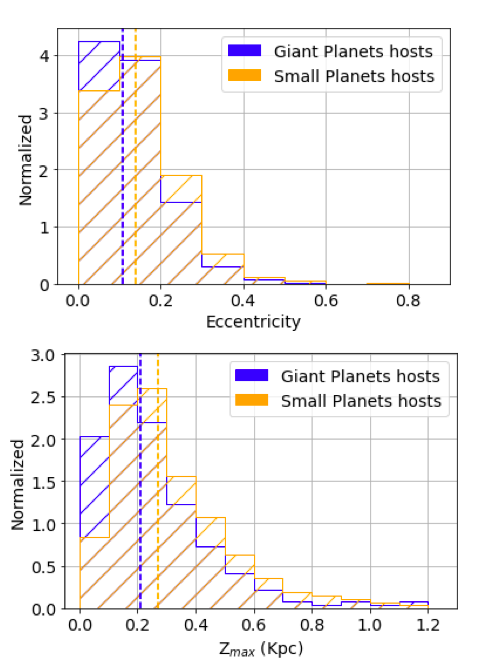}
\caption{Galactic orbital parameters for the small and giant planet-hosting stars. \textbf{Top:} Eccentricity distribution for the planet hosting stars (binned at 0.1 dex). \textbf{Bottom:} $Z_{max}$ distribution for the different population of exoplanet hosting stars (\textbf{binned at 0.1 Kpc}). The vertical lines represent the median of the distribution for the small and giant planet hosts.}
\label{f3}
\end{figure}

\begin{table*}[]
\centering
\caption{\label{t3} Key parameters of exoplanet hosting stars used or estimated in this study.}
\begin{tabular}{llllllll}
\hline\hline
TIC ID & hostname & Method & Planet & Pl- Mass(M$_{J}$) & RA & DEC & Parallax (mas) \\
\hline
TIC 328465904 & CD Cet & Radial Velocity & CD Cet b & 0.01243 & 48.3530155 & 4.7751881 & 116.267814433972  \\
TIC 380966347 & HD 14787 & Radial Velocity & HD 14787 b & 1.121 & 35.8085099 & 10.8367972 & 8.45391699045641  \\
TIC 435339847 & K2-77 & Transit & K2-77 b & 1.9 & 55.228521 & 12.572448 & 7.08178354133806  \\
TIC 435339558 & K2-79 & Transit & K2-79 b & 0.0415 & 55.2559307 & 13.5191871 & 3.8412449465656  \\
TIC 242961495 & K2-80 & Transit & K2-80 b & 0.0148 & 59.037486 & 13.5590288 & 5.02001944435433  \\
TIC 242961495 & K2-80 & Transit & K2-80 c & 0.00869 & 59.037486 & 13.5590288 & 5.02001944435433 \\
... & & & & & & & \\
\hline
\end{tabular}
%\end{table}

\tablecomments{The entire table is available in machine-readable format. For simplicity, only the first 6 rows and 8 columns are shown here.}
\end{table*}
%\footnote{https://www.cosmos.esa.int/web/gaia-users/archive/gdr3-documentation}
%The abundances were then calibrated as a polynomial function of $log\:g$, similar to metallicity.
\section{The sample}
\label{s2}
For this study, we used the confirmed list of exoplanetary systems from  NASA exoplanet archive \citep{2013PASP..125..989A,https://doi.org/10.26133/nea12} and cross-matched it with the latest Gaia data release DR3 to obtain the stellar data for the planet-hosting stars. We first employed the Astronomical Data Query Language (ADQL) to identify the GAIA DR3 source IDs associated with our exoplanet-hosting stars. Subsequently, we utilized ADQL to extract the corresponding data from the Astrophysical parameters table. In cases where multiple matches were found, we manually verified the G-Band magnitude quoted in the NASA exoplanet archive and selected the closest match to the search result in the GAIA DR3 dataset. Additionally, for the purpose of validation, we used TOPCAT to perform Ra-Dec cross-matching with a search radius\footnote{Initially, we used a larger search radius and found that most of the planet-hosting stars can be extracted with a search radius of $3''$. For cases where we were not able to obtain the matches, we increased the search radius up to $15''$ and also checked for the G-band magnitude to confirm if the target is indeed a planet-hosting star.} of $3''$ and obtained identical results to those extracted using ADQL.

In the case of GAIA, the spectroscopic data is obtained from the Radial Velocity Spectrometer(RVS) instrument on board GAIA, which is a medium-resolution spectrograph. The data products from the RVS spectra are listed in the \textit{Astrophysical parameters} table. Further, each parameter is associated with a \textit{quality flag}\footnote{For more details on \textit{quality flag}, see Table 2 of \cite{2022arXiv220605541R}} indicating quality of the data. For the analysis presented in this paper, most of our samples are associated with the best-quality flag (0 in this case) and we excluded the stars with low-quality data flags (9 in this case).  
Our primary sample, therefore, consists of 2611 planet-hosting stars (accounting for 3553 planetary companions) for which the radial velocity data was available from the GAIA archive. The sample extracted from GAIA as well as the important parameters derived in this paper are listed in Table~\ref{t3}. Figure~\ref{f1} shows the distribution of these stars in Mollweide projection, while Figure~\ref{plmass} shows the distribution of planet mass whose sources are listed in the GAIA archive.  

Since the original sample contained  many evolved stars, notably giants and subgiants, we  restricted our analysis to the main sequence stars. The reason is that it is difficult to account for the NLTE and other evolutionary effects which can alter the surface abundances of the evolved stars \citep{2022AJ....164...60S}. We followed the procedure of \cite{2018ApJS..237...38B} to exclude the evolved stars using $T_{eff}$ and $\log g $ cutoff. Further, we included only those host stars where the companion mass $<13 M_{J}$ and also to avoid biases from possible mixing, we also excluded stars with multi-planetary systems containing a combination of small and giant planets. After applying these filters, our final sample was trimmed to 971 stars with 1309 planets for which the spectroscopy data is available and 2130 stars with 2861 planets for which astrometric data is available. Further, after estimating the stellar ages (more details in section~\ref{s33}) we included only those stars whose uncertainties are $<$ their main-sequence lifetime as suggested by \cite{2004MNRAS.351..487P}. Also, we excluded the lower main-sequence stars from our age sample (T$_{eff}<$4400K), as the isochrone ages for the lower main-sequence stars are not very accurate given the large uncertainties. Thus after curation, we analysed the ages of 806 stars hosting 1071 planets. In the sample described above, about $\sim 83\%$ of stars in the astrometrically curated sample belong to transit surveys (mostly Kepler) and $\sim 17\%$ belong to different RV surveys.  In the spectroscopic sample, $\sim 64\%$ stars are from transit discoveries while the remaining come from the RV detections. Additionally, the spectroscopic sample is a subset of the astrometric sample, meaning the astrometric data is available for all stars belonging to the spectroscopic sample. 
%The entire data used in this paper is listed in Table~\ref{t3}.

\section{Results}
\label{s3}
We used different proxies of stellar ages from spectroscopic, photometric and astrometric data from the GAIA DR3 to analyze the confirmed exoplanet population. Below, we present our results obtained from the GAIA DR3 data in the context of planet formation.
\subsection{Spectroscopic analysis of the planet hosts stars}
The Gaia DR3 provides a significantly large sample of stars whose spectroscopic parameters are determined homogeneously. The GSP-Spec module \citep{2022arXiv220605541R} does the spectroscopic processing using the combined Radial Velocity Spectrometer (RVS) spectra of single stars to calculate stellar chemo-physical characteristics. The RVS covers a spectral range of 846-870~nm and has a resolution of R $\sim$ 11500 \citep{2018A&A...616A...5C}. The GSP-Spec module estimates the stellar atmospheric parameters ($T_{\textrm{eff}}$, $\log g$, [M/H]\footnote{Here [M/H] is defined as the total metal content of the star.}) and the abundances of 13 chemical species (N, Mg, Si, S, Ca, Ti, Cr, Fe I, Fe II, Ni, Zr, Ce and Nd). The stellar atmospheric parameters are estimated using the Matisse GAUGUIN algorithm and artificial neural network (ANN) \citep{2016A&A...585A..93R,2022arXiv220605541R}. However, the abundances are obtained solely from the Matisse GAUGUIN algorithm using Gaussian fitting methods \citep{2021A&A...654A.116Z,2022arXiv220605541R}. For the analysis presented in this paper, we used the stellar parameters and abundances from the Matisse GAUGUIN algorithm. 

Since GAIA spectroscopic data from the GSP-Spec module suffers from estimation biases \citep{2022arXiv220605541R}, we used the HARPS-GTO sample (a high-resolution sample of 1111 stars targeted mainly with the goal of detecting planets by radial velocity) for calibration \citep{2003Msngr.114...20M,2010A&A...512A..48L,2011A&A...526A.112S}. After taking care of the calibration and possible estimation biases, as discussed in Appendix A, we investigated the host star metallicities and [Mg/Fe] (a proxy for overall $\alpha$ abundances) in the GAIA archive as a function of planet mass. We chose [Mg/Fe], since we wanted to investigate the ratio of the abundances of elements produced from Type II supernovae (Mg) to Type I supernovae (Fe) and since the major production site for Mg is Type II supernovae, it is the strongest tracer for the overall $\alpha$ abundance in a star \citep{2020ApJ...900..179K}. We used the planet mass from the NASA exoplanet archive \citep{2013PASP..125..989A,https://doi.org/10.26133/nea12}\footnote{For the planets detected by transits, we used the planet mass-radius relationship from \citep{2017ApJ...834...17C}. For the planets detected by RV, we used the minimum mass (M.sini) as listed in the NASA exoplanet archive.} and binned the data appropriately in terms of planet mass depending on the number of stars in each bin, with four bins for small planet-hosting stars ($M_{P}<$0.3$M_{J}$), two for giants ( 0.3$M_{J}\leq M_{P}\leq 4M_{J}$), and one for super-Jupiters ($M_{P}>$ 4 $M_{J}$).  We found that the host star metallicity increases as a function of planet mass with a turn-around after $\sim$ 4 $M_{J}$ as seen in Figure~\ref{f2}. Although several studies have shown similar results \citep{fis05,2008ASPC..384..292V,nar18,2021AJ....161..114S}, they were mostly limited to either small samples or inhomogeneous measurements of metallicities. Here, in this paper, we could reproduce these results for a large number of exoplanet-hosting stars using the data from the RVS spectra from the GAIA DR3. 

We also find that, for the $\alpha$-element abundances ([Mg/Fe]), there is a decreasing trend with planet mass, as seen in Figure~\ref{f2}. For comparison with $\alpha$ element, we used only Fe abundances, as abundances of only two other Fe-peak elements (Ni and Cr) were available and Fe is estimated with much better precision compared to Ni and Cr in the GAIA GSP-Spec module. Since [Fe/H] and $\alpha$-enhancement are proxies for ages for a population of stars \citep{2022AJ....164...60S,2019A&A...624A..78D}, the decline in $\alpha$-abundances, together with enhancement of [Fe/H] indicate that giant planets are preferentially hosted by younger stars while the stars having small planetary companions have a wider spread in the age.
\begin{table}
\caption{\label{t1}Comparison of small and giant planet-hosting stars in terms of their galactic parameters and ages determined from the isochrone fitting.}
\centering
\begin{tabular}{lccc}
\hline\hline
&Small-planet& Giant-planet& p-Value\\
\hline
\\
Thin disk & 1464 & 579 & $>$0.05\\
Thick disk & 29 & 7 & $>$0.05 \\
$Z_{max}$ (kpc) &0.27$\pm$0.12 & 0.21$\pm$0.09 &  10$^{-4}$ \\
Eccentricity &0.14$\pm$0.07 & 0.11$\pm$0.06 & 10$^{-5}$\\

$\nu_{pec}$ (km/s)&42.79$\pm$ 0.35 & 33.19$\pm$0.18 &  10$^{-6}$\\
$\sigma_{tot}$ (km/s) & 53.70$\pm$0.41 & 42.89$\pm$0.41 &--\\
Ages (Gyr) &4.07$\pm$3.23 & 3.17$\pm$2.67& 10$^{-6}$\\
\hline
\end{tabular}
\tablecomments{Errors represents the 1$\sigma$ spread in the corresponding distribution of the parameters, except for $\nu_{pec}$ and $\sigma_{tot}$ where the  1$\sigma$ spread is obtained from Monte Carlo method by taking into account for the uncertainties in U, V and W (see text for more details). The p-value represents the probability of two samples belonging to the same distribution using the Anderson-Darling test. \\
}
\end{table}
\subsection{Kinematic analysis of exoplanet hosting stars}
\label{s32}
The kinematic analysis of stars entails tracking the past motions of a group of stars to determine when they were physically closest, which is thought to be the period of their formation. In this case, stellar parameter estimation such as the galactic space velocities (U, V, W) and orbital parameters (eccentricity and $Z_{max}$) is based on minimal assumptions and does not need stellar modelling but high-quality astrometry and radial velocities measurements. In our case, we used the radial velocity and proper motion data from the GAIA DR3 data to compute the galactic space velocities \citep{1987AJ.....93..864J,2020AJ....159..166U}.
Further, we used galpy \citep{2015ApJS..216...29B} to compute the stellar orbital parameters (eccentricity and $Z_{max}$) and used the solar motion ($U_{\odot}$, $V_{\odot}$, $W_{\odot}$) = (11.1, 12.24, 7.25) kms${^{-1}}$from \cite{2010MNRAS.403.1829S} as a reference. We analysed a sample of 2130 stars and found that the stars hosting small planets have higher median  eccentricity and $Z_{max}$\footnote{$Z_{max}$ is integral of motion that tells us the maximum height above or below the galactic plane on the disk that a star travels.} compared to giant planet-hosting stars, as seen clearly in Figure~\ref{f3}. 
\begin{figure}[t]
\centering
\includegraphics[width=1\columnwidth]{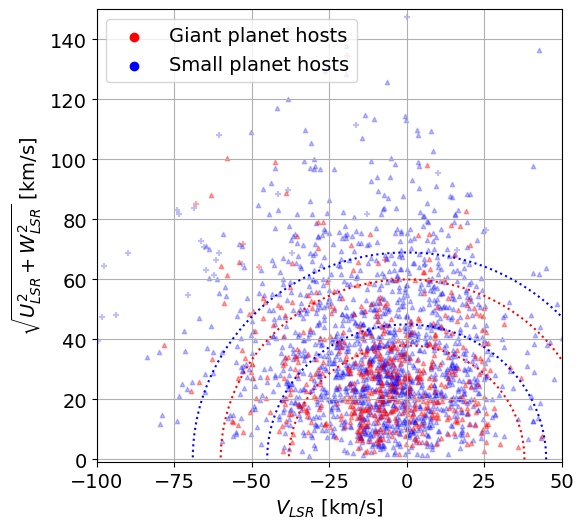}
\caption{Toomre diagram for the current sample of planet-hosting stars. The blue and red circles represent the locus of $\nu_{pec}$ for the small and giant planet hosts. The area enclosed by inner circles has $\sim50\%$ of stars population while the outer circles capture $\sim80\%$ of the population in each category. }
\label{f4}
\end{figure}

\begin{figure*}[t]
\centering
\includegraphics[width=2\columnwidth]{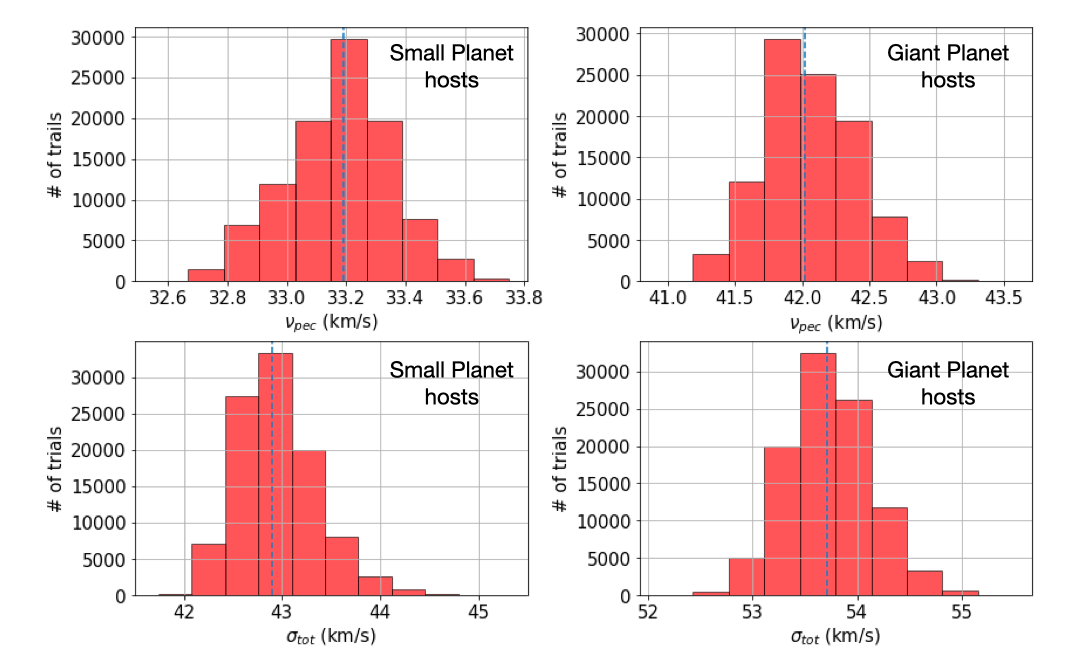}
\caption{Distribution of $\nu_{pec}$ and $\sigma_{tot}$ for the planet-hosting stars obtained from Monte Carlo simulation. The vertical dashed lines represent the median of the distribution.}
\label{pecsig}
\end{figure*}

Further, peculiar velocity ($\nu_{pec}$)\footnote{$\nu_{pec}^2$ = $U^{2}_{LSR}$+$V^{2}_{LSR}$+$W^{2}_{LSR}$;  It is represented by the radius drawn from the origin in the Toomre diagram} and total velocity dispersion ($\sigma_{tot}$)\footnote{ $\sigma_{tot}^2$ = $\sigma_{U}^2 + \sigma_{V}^2 + \sigma_{W}^2$ } distribution have notable difference for small and giant planet-hosting stars as shown by the red and blue circles in Figure~\ref{f4}. We find that the scatter in the $\nu_{pec}$ is much more significant for small planet hosts than for giant planet-hosting stars. In the case of small planet hosts, e.g., 50 and 80 percent of the population lies at the velocity radius of 46~km/s and 69~km/s, compared to 38~km/s and 60~km/s for giant planet hosts. Since the age for an ensemble of stars increases radially from the origin, with the thin disk (younger population) stars having low $\nu_{pec}$ and extending to thick disk and halo stars (older population) having higher $\nu_{pec}$ \citep{2006MNRAS.367.1329R,2011A&A...530A.138C}. The clustering of the giant planet hosts  around the origin of the Toomre diagram (Figure~\ref{f4}) indicates that they belong to a statistically  younger population of stars compared to stars hosting small planets, which show a larger spread in $\nu_{pec}$ (and $\sigma_{tot}$). Considering the uncertainties associated with Gaia proper motion, RV (radial velocity), and parallaxes, we conducted an additional assessment to investigate the impact of these uncertainties on the estimation of $\nu_{pec}$ and $\sigma_{tot}$. To account for uncertainties in the space motion of stars, we calculated the error in U, V, and W using the relationships described in equation 2 of \cite{1987AJ.....93..864J}. The median uncertainties in U, V, and W were found to be 0.16, 0.49, and 0.17 km/s, respectively. To determine if these uncertainties affect the analysis presented in this paper, we performed a Monte Carlo simulation where each space velocity component U, V, and W  of a star is randomly generated from a Gaussian distribution with the mean and standard deviation obtained as described above. We then calculated $\nu_{pec}$ using these random realisations of U, V, and W for both small and giant planet-hosting stars. This process was repeated 100,000 times and we find that the 1$\sigma$ spread in $\nu_{pec}$ for small and giant planet hosts is 0.35 and 0.18 km/s, which is relatively small (Figure~\ref{pecsig}, top row) compared to the absolute difference in the $\nu_{pec}$ between small and giant planets ($\sim$ 10 km/s), suggesting that the uncertainties in the Gaia astrometric parameters do not significantly affect the analysis presented in this paper. We also conducted a similar analysis for $\sigma_{tot}$ as shown in the bottom row of Figure~\ref{pecsig}. We have also noted the median and spread obtained from the Monte Carlo analysis for $\nu_{pec}$ and $\sigma_{tot}$ in Table~\ref{t2}.

Further, we classified the likelihood of each star belonging to the thin disk, thick disk, or halo using the approach adopted by \cite{2006MNRAS.367.1329R}. Therein, the parent sample is considered to be a mixture of the three populations, and it is assumed that every population has a Gaussian random distribution of  velocities for each component \citep{2006MNRAS.367.1329R,2011A&A...535L..11A}. By assigning a probability threshold of 70$\%$ for a star to belong to a particular population, we find $\sim98\%$ of the planet-hosting stars belong to a thin disk population (Table~\ref{t1}). We also find that the sample of stars hosting small and giant planets cannot be manifested in terms of thin vs thick disk population. In terms of galactic orbital parameters, we find that on average, stars hosting small planets have higher median eccentricity and $Z_{max}$ compared to giant planet-hosting stars. We also performed an Anderson-Darling test (AD) and found that the difference is significant for both galactic space velocities and orbital parameters (Table~\ref{t1}), suggesting that small and giant planet-hosting stars likely belong to different populations. 

Several studies \citep[e.g.][]{2003ScChA..46....1C,2011A&A...530A.138C,2018MNRAS.477.5612W,2019ApJ...883..177N,2022MNRAS.510.3449B} have indicated that the higher values of $Z_{max}$, eccentricity and $\sigma_{tot}$ are a proxy for older stars. For instance, \cite{2018MNRAS.477.5612W} found that eccentricity differ by $\sim 0.05$ and $z_{max}$ by $\sim$~0.04 for the young ($\le $ 3Gyr) and the old ($\ge8$~Gyr) stars. For exoplanet hosting stars, using a limited sample (135 stars) of Neptune, super-earth and Jupiter hosts, \cite{2012A&A...543A..89A} have also shown that Jupiter hosting stars have lower median eccentricities and $Z_{max}$ compared to stars hosting Neptunes (see table 3 of \cite{2012A&A...543A..89A}). In our study, this is validated for a larger sample of exoplanet-hosting stars using the astrometric and radial velocity data from GAIA. We note that the distribution of $Z_{max}$, eccentricity, $\nu_{pec}$ and $\sigma_{tot}$ are statistically different for stars hosting small and giant planets. For comparison, these parameters along with the p-value are listed in Table~\ref{t1}.
%Our results indicate that  stars hosting giant planets belong to younger populations while stars hosting small planets belong to both younger and older populations.
% In this we analysed a sample of 1380 planets hosted by 1030 stars.
% error in (U,V,W ) =(0.0792342987015443, 0.0861693363088793, 0.0908793744832608)

\subsection{Ages of planet-hosting stars}
\label{s33}
Yet another way to distinguish the parent stars of small and giant planets is to directly estimate their ages. Asteroseismology is the only technique that can determine a star's age with uncertainty as low as 11$\%$ \citep{2019A&A...622A.130B}. However, it needs high-cadence photometric observations of stars spread over a long observation window which is available only for a few hundred targets. Additionally, it only applies to stars hotter than about spectral type K, as cooler stars generally do not show oscillations, which are necessary for determining the ages using asteroseismology  \citep{2018haex.bookE.184C,2015MNRAS.452.2127S}. The isochrone fitting approach, in which the ages are determined by positioning the star in the HR diagram, is another well-known and often employed technique. However, isochrone fitting can have large age uncertainties, typically up to 20$\%$ or more  for the main-sequence stars \citep{2022ApJ...927...31T}. 

\begin{figure}[t]
\centering
\includegraphics[width=1\columnwidth]{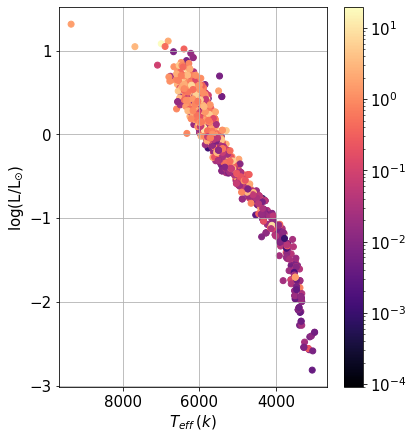}
\caption{Hertzsprung-Russell Diagram for Stars with Determined Ages in this Study. The color bar on the left indicates the planet mass in units of $M_j$.}
\label{hrd}
\end{figure}

\begin{figure}[t]
\centering
\includegraphics[width=1\columnwidth]{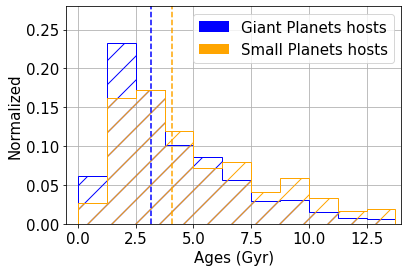}
\caption{Age distribution for a sample of 807 planet-hosting stars from the MIST isochrone models (binned at 1.25 Gyr). The vertical lines represent the median age.}
\label{Age}
\end{figure}

Nevertheless, to independently check the age trends in our sample, we used isochrone models from the Modules for Experiments in Stellar Astrophysics  (MESA) \citep{2011ApJS..192....3P,2013ApJS..208....4P,2015ApJS..220...15P,2018ApJS..234...34P} isochrones \& Stellar Tracks (MIST) \citep{2016ApJ...823..102C,2016ApJS..222....8D} using the python based isoclassify \footnote{https://github.com/danxhuber/isoclassify} \citep{2017ApJ...844..102H,2020AJ....159..280B}. For the input parameters, we used GAIA spectroscopic parameters that are listed in the \textit{Astrophysical parameters} table (calibrated as outlined in Appendix A), together with parallaxes and G-band magnitudes taken from the Gaia DR3 data. Typical uncertainties in stellar atmospheric parameters T$_{eff}$, $\log g$ and [Fe/H] are assumed to be  100~K, 0.1~dex and 0.1~dex, respectively. The H-R diagram for the stars whose ages are estimated in this paper is shown in Figure~\ref{hrd}. After applying the cutoffs as described in section~\ref{s2}, the distribution for the stellar ages for our sample is shown in Figure~\ref{Age}. We find that the median age of giant planet-hosting stars to be $\sim 3.17$~Gyr and for the small planet hosts to be $\sim4.07$~Gyr (Table~\ref{t1}). We performed a AD test which yielded a small p-value ($p = 1.22 \times 10^{-5}$), indicating that the two datasets differ significantly in terms of their underlying distributions. However, given the large uncertainties in the individual age estimates using isochrone fitting, the AD test alone may not be a reliable predictor of the statistical significance of the age difference between the two distributions. To assess whether the population-level difference is statistically significant we need to account for the individual age uncertainties associated with isochrone fitting.  For that, we again performed a Monte Carlo experiment similar to the one described in section~\ref{s32}. For each star, we randomly draw the age from the Gaussian distribution whose mean and sigmas are estimated from the isochrone modelling. Repeating over the entire sample, the MC age distribution was obtained for stars with small and giant planet-hosting stars. We then compare two populations  to obtain the p-value using the AD test. Once the p-value is noted, we repeated the analysis 1,00,000 times. From the assemblage of p-values, we find that  about $99\%$ of the time, the p-value was smaller than 0.05. This numerical experiment clearly indicates that the population-level differences in  isochrone ages shown in Figure~\ref{Age} are statistically significant.

\section{Discussions}
\label{s4}
Several galactic orbital ($\sigma_{tot}$, eccentricity and $Z_{max}$) and spectroscopic ([Fe/H] and [$\alpha$/Fe]) parameters are proxy to stellar ages. To study different exoplanet populations and their formation timeline, we investigated the ages of their host stars. Our analysis shows that small planet-hosting stars have higher $\sigma_{tot}$, eccentricity, $Z_{max}$ and [$\alpha$/Fe] and lower [Fe/H] compared to stars hosting giant planets. Since higher $\sigma_{tot}$, eccentricity, $Z_{max}$  and [$\alpha$/Fe] are indicators of older population \citep{2003ScChA..46....1C,2011A&A...530A.138C,2018MNRAS.477.5612W,2019ApJ...883..177N,2022MNRAS.510.3449B}, we find that the small planet-hosting stars are statistically older compared to their giant planet hosts. Further, to validate this, we used the isochrones fitting technique to estimate the stellar ages, using MIST isochrone grids, and we arrived at similar conclusions. While the majority of our planet-hosting stars primarily belong to the thin disk population and exhibit a predominance of higher metallicity (refer to Figure A1 of \cite{2022AJ....164...60S}), we conducted additional investigations to determine what extent the stars hosting small and giant planets are younger. This analysis involved controlling for the correlation between planet mass and stellar metallicity, considering that stellar ages are directly influenced by various stellar properties, including mass and radius. For instance, for a controlled stellar sample with the criteria of -0.2$<$[Fe/H]$<$0.4 and 0.7R$_{sun}<$R$_{star}<$1.3R$_{sun}$, we examined the extent to which the histogram offsets persisted. Notably, although the offsets were still observable, they exhibited a decrease and the histogram peaks shifted towards younger ages. This is expected since we selectively removed stars from specific age groups (older in this case) within the sample. We also repeated this analysis for other combinations as well and found similar trends. The fact that young, metal-rich stars have a preference for hosting giant planets aligns with the natural progression of the chemical evolution of the galaxy and it is more challenging for giant planets to form around older, metal-poor stars.

Radial velocity (RV) and transit detection are two of the most popular techniques used to detect exoplanets. However, these techniques have inherent biases that can impact our understanding of exoplanet populations \citep{2022AJ....164...60S}. 
The radial velocity technique, e.g., can detect massive planets that are close to their host star and have intermediate orbital periods (up to $\sim$ 10 AU). However stellar activity and line-broadening mechanisms reduce RV precision, and therefore very active and fast-rotating stars are usually excluded from the RV surveys. The transit method on the other hand is more sensitive to short orbital period planets (mostly below $\sim$ 1 AU) whose orbits are favourably aligned along the observer's line of sight. Both of these methods have their own detection biases and therefore lead to a non-representative sample of the true exoplanet population in the galaxy. For instance, younger stars have large variability, therefore, finding smaller planets around stars is more challenging \citep{2016MNRAS.459.3565V}.
%However, the results presented in this paper is  unlikely due to any known selection or detection bias. 

\begin{figure}[t]
\centering
\includegraphics[width=1\columnwidth]{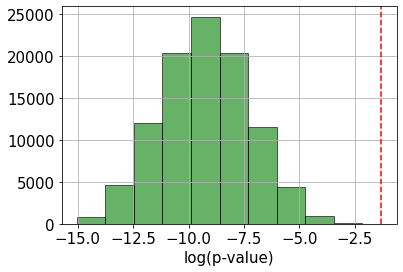}
\caption{Distribution of p-values from the 1,00,000 AD test performed using monte carlo method. The black dashed line represents the p = 0.05, below which two distributions are considered to be statistically different.}
\label{pval}
\end{figure}

It is possible that some small planets might have missed detection around young stars due to sensitivity limitations. However, different age proxies used in this work indicate that older stars have fewer giant planets compared to younger stars. The fact that giant planets are easier to detect irrespective of detection technique or the age of the star indicates that the overall occurrence rate of giant planets is much lesser around old stars (also shown from isochrone ages using the Kepler sample in Swastik et al in prep) and suggest that giant planets may have started forming late in the galaxy.  

Stars and planets both originate from the same molecular cloud within the interstellar medium (ISM), and the metal content in the ISM is a crucial factor in the formation of planets. Our study reveals a significant finding that stars hosting giant planets are statistically associated with a younger population. For the formation of a giant planet, a core of  $\sim$10 M$_{\earth}$ must be formed within a relatively short timeframe of around 10 million years \citep{1996Icar..124...62P} before the dissipation of gas in the protoplanetary disk. This core primarily consists of refractory elements, including both $\alpha$-elements (such as Mg, Si, Ca, etc.) and Fe-peak elements (such as Fe, Ni, etc.). During the demise of the first stars as core-collapse supernovae (SNe II), the interstellar medium (ISM) became enriched with $\alpha$-elements. However, during the early stages of the Milky Way's existence, the ISM lacked sufficient enrichment in heavy elements, particularly Fe-peak elements. This limitation hindered the core-accretion process necessary for the formation of giant planets \citep{Rice03,2007ApJ...662.1282M,2022arXiv220309759D}. With the gradual enrichment of the ISM through Type~Ia supernovae (SNe Ia), the availability of more Fe-peak elements, such as Fe, Ni, Cr, Mn, etc., facilitated the onset of giant planet formation \citep{mat89,ali01,mat09,2020ApJ...900..179K,2022AJ....164...60S}. Therefore, the scarcity of giant planets around older stars and  the widely observed  planet-mass and stellar-metallicity correlation can be understood as a natural outcome of the galactic chemical evolution of the Milky Way. Seen that way, the temporal offset between the formation of small and giant planets will  also be consistent with  chemo-kinematic trends of planet-hosting stars and the mass-metallicity relationship.

\section{Conclusions}
\label{s5}
The properties of exoplanets are closely related to the traits of their stellar hosts. In this work, we studied planet-hosting stars' chemical abundances, kinematics, and their ages. We used the GAIA DR3 data, for which the stellar parameters are available for the large number of exoplanet-hosting stars whose parameters have been estimated uniformly. We analyzed the astrometric, photometric and spectroscopic data from the GAIA DR3. Here, we present a brief summary of our analysis :
\begin{itemize}
    \item Using the GAIA spectroscopic metallicities and abundances from the RVS spectra, we find that the host stars of giant planets are metal-rich and $\alpha$-poor compared to small planet-hosting stars. This finding indicates that host stars of giant planets belong to a younger population of stars which started forming in the later stages of the galaxy after the enrichment of ISM with Fe-peak elements.
    \item We find that most of our planet-hosting stars belong to the thin disk population, indicating that the overall sample of exoplanets-hosting stars belongs to the younger generations. For the galactic space velocities and orbital parameters, we find that host stars of small and Jupiter-like planets belong to a separate population. We also find that small planet-hosting stars have higher $Z_{max}$ and eccentricities (which is a signature for older stars) compared to giant planet-hosting stars.
    \item By using the MIST isochrone models, we were able to estimate the approximate ages of the stars which host exoplanets. Our analysis reveals that stars which host giant planets are likely to be younger than those which host small planets, despite the fact that there are considerable uncertainties in the age estimates that are obtained from isochrones.
\end{itemize}
The present observations using the latest GAIA DR3 data suggest that the giant planets started forming at the later stages of the GCE evolution when the ISM was sufficiently enriched with Fe-peak elements by Type Ia supernovae, which happened around $\lesssim$ 6 Gyr. The enrichment of ISM is necessary to form the core of the giant planets faster before the dissipation timescale of the gas in the protoplanetary disk. Our results are also consistent with the core-accretion theory of planet formation \citep{1996Icar..124...62P, 2007ApJ...662.1282M,2016SSRv..205...41B,2018MNRAS.480.2206O,2022arXiv220309759D}.
Future missions consisting of astrometry, photometry and spectroscopic investigations should focus on a larger sample of exoplanet-hosting stars, measuring their chemical abundances and astrometric parameters uniformly and more precisely to support these findings.  
\software{Numpy \citep{2020Natur.585..357H}, Topcat \citep{2005ASPC..347...29T}, Astropy \citep{astropy:2013}, Scikit-learn \citep{scikit-learn}, Matplotlib \citep{4160265}, Scipy \citep{2020NatMe..17..261V} }

\begin{acknowledgements}
This work has made use of data from (a) the NASA Exoplanet Archive, which is run by the California Institute of Technology under an Exoplanet Exploration Program contract with NASA,  (b) the  GAIA space mission by the European Space Agency (ESA)  (the data was processed by the {\it Gaia} Data Processing and Analysis Consortium (DPAC), and funding for the DPAC has been provided by national institutions, in particular, the institutions participating in the {\it Gaia} Multilateral Agreement) and  (c) the exoplanet.eu database which compiles and organizes data from various sources, including ground-based and space-based telescopes. We gratefully thank the anonymous referee for the insightful review and comments that helped us improve the paper. Additionally, C. Swastik would like to thank Luca Casagrande, Partha Pratim Goswami and Sioree Ansar for numerous discussions on stellar kinematics.
\end{acknowledgements}

\begin{appendix}
\label{a1}
\section{Calibration of Metallicities and abundances}
\begin{figure}[b!]
\centering
\includegraphics[width=0.5\columnwidth]{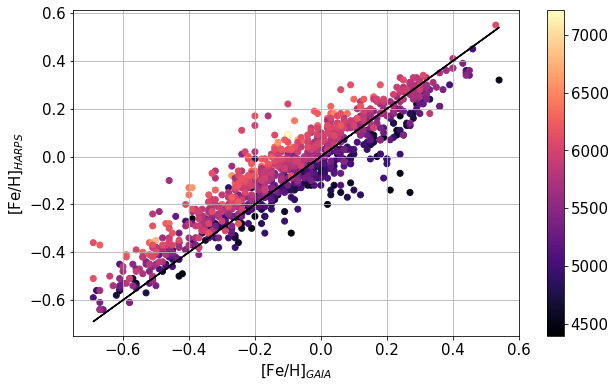}
\caption{Comparison of the [Fe/H] values from GAIA GSP-Spec module and the HARPS sample. The black  line is the x=y line and the color bar on the right represents the effective temperature of the star.}
\label{fa1}
\end{figure}
\begin{table}[h!]
\caption{\label{t2} Differences in median offset and robust sigma between GAIA GSP-Spec and individual surveys.}
\centering
\begin{tabular}{l|ccc|cc}
\hline\hline
& T$_{\mbox{eff}}$& $\mbox{log(g)}$& [M/H]& $\mbox{log(g)}_{calibration}$& [M/H]$_{calibration}$\\
\hline
% & &  & & (-0.18, 0.24) & (-0.07, 0.09)\\
RAVE-DR6 & (-12; 93) & (-0.28; 0.19) & (-0.05; 0.11)& (-0.003; 0.18) &  (-0.05; 0.09)\\
APOGEE-DR17 &(-32; 86) &  (-0.32; 0.17) & (0.04; 0.12)&(-0.005; 0.15) & (0.06; 0.12) \\
GALAH-DR3 & (20; 87) &  (-0.26; 0.21) & (0.01; 0.10)& (0.003; 0.18) &  (-0.001; 0.10)\\
\hline
HARPS-GTO (current work using equation~\ref{eq1})) & (-12, 97) & (-0.24, 0.30) & (-0.04, 0.08)& (-0.05, 0.25) & (0.001, 0.07)\\
HARPS-GTO (with GAIA proposed polynomial) & & & & (-0.18, 0.24) & (-0.07, 0.09) \\
\hline
\end{tabular}
\tablecomments{ The first three rows are taken from Table D1 of \cite{2022arXiv220605541R}. The values in the parenthesis are median offset, followed by robust sigma (standard deviation obtained by removing outliers) computed from the residuals}. The last two rows are for HARPS-GTO data with the calibration results from equation~\ref{eq1}, and GAIA proposed polynomials.
\end{table}

\begin{figure}[h!]
\centering
\includegraphics[width=0.5\columnwidth]{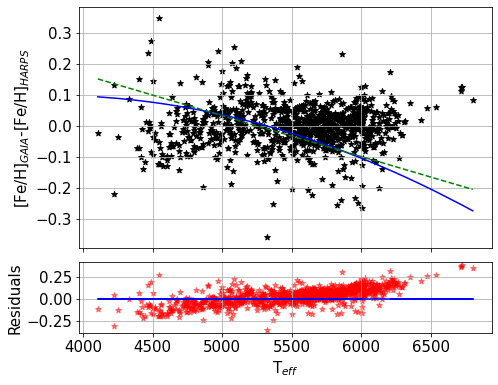}
\caption{Comparison of $\delta$[Fe/H] =  [Fe/H]$_{GAIA}$-[Fe/H]$_{HARPS}$ as a function of T$_{eff}$. A quadratic polynomial best describes the trends of $\delta$[Fe/H] with T$_{eff}$. The blue is the best-fit polynomial with the coefficients ( p0, p1, p2) = (-3.96e-08,  2.97e-04, -4.57e-01) for the top figure, while the bottom figure shows the residuals in the approximation of the trend by the above polynomial. The green dashed line shows the linear fit for the data.}
\label{fa2}
\end{figure}

Although GAIA-DR3 provides the homogeneous estimation of stellar parameters for the largest number of stars to date, it requires several calibrations and filtering for any rigorous scientific study. For instance, the data from the GAIA-GSP spec module suffers from systematic \citep{2022arXiv220605541R} and hence one needs to account for such biases to perform any meaningful analysis on the data. Although some calibrations have been already been proposed for GAIA data using three major ground-based surveys: APOGEE-DR17 \citep{abd}, RAVE-DR6 \citep{Steinmetz_2020}, and GALAH-DR3 \citep{2021MNRAS.506..150B}, we still find a significant offset and scatter in the calibrated data as shown in Table~\ref{t2} (last two rows) when compared to high-resolution and high SNR spectroscopic data from the HARPS-GTO sample \citep{2003Msngr.114...20M,2010A&A...512A..48L,2011A&A...526A.112S}. Also, the calibration polynomials are established for stars with a wide range of atmospheric parameters in $\log g$, and T$_{eff}$ and thus, we decided to use our own tailored calibration for our sample of planet-hosting stars. Therefore, we use the HARPS-GTO sample \citep{2003Msngr.114...20M,2010A&A...512A..48L,2011A&A...526A.112S}, which is a survey of 1111 stars that were chosen to detect planets by radial velocity and also have a similar range of atmospheric parameters as that of the planet-hosting stars used in this paper. Although the wavelength coverage of HARPS (378-691 nm) and GAIA (846-870 nm) is different, it will not affect the estimate of stellar atmospheric parameters as there are sufficient Fe-lines to estimate the metallicities. We cross-matched our sample and found 932 common stars between the HARPS and the GAIA samples. Upon comparing the metallicities of the stars from GAIA and HARPS, as shown in Figure~\ref{fa1}, we find that the distribution about the $x=y$ line is not symmetric. In addition, we also find a temperature gradient with [Fe/H], where the [Fe/H] is underestimated for hotter stars ($\geq$ 5500K) and overestimated for cooler stars ($\le$ 5500K). Therefore, to account for the temperature dependence, we  analysed the $\delta$[Fe/H] =  [Fe/H]$_{GAIA}$-[Fe/H]$_{HARPS}$ as a function of T$_{eff}$. The relationship between the $\delta$[Fe/H] and $T_{eff}$ can be best described by a quadratic polynomial (Figure~\ref{fa2}) given below: 

\begin{equation} \label{eq1}
  \delta \textrm{[Fe/H]} = \textrm{[Fe/H]}_{GAIA} - \textrm{[Fe/H]}_{HARPS} = \sum_{i=0}^{2}p_{i}. T_{eff}^i
\end{equation}

Thus, for a given $T_{eff}$ we compute the $\delta$[Fe/H] using Eq.~\ref{eq1} to estimate the offset in [Fe/H]. We then calibrated our data using Eq.~\ref{eq1} and then checked for any remaining offset in the HARPS-GTO and GAIA calibrated data using our calibrations. We find no significant offset between GAIA calibrated and HARPS-GTO data as listed in Table~\ref{t2}. We then applied the same correction to the sample of planet-hosting stars to correct the estimation bias and then used those calibration values to study the correlation trends.

 \begin{figure}[h!]
\centering
\includegraphics[width=0.5\columnwidth]{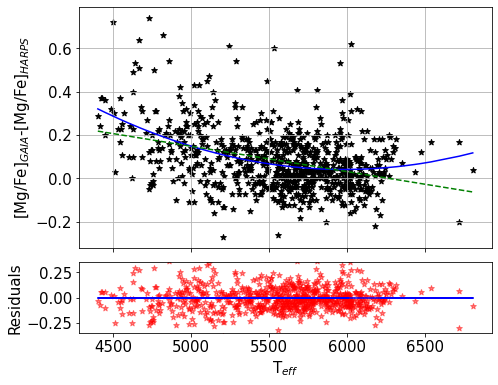}
\caption{Variation of $\delta$[Mg/Fe] =  [Mg/Fe]$_{GAIA}$-[Mg/Fe]$_{HARPS}$ as a function of effective temperature. A quadratic polynomial best describes the trends of $\delta$[Mg/Fe] with T$_{eff}$. The best-fit polynomial is represented by the blue line in the top figure, by the coefficients ( p0, p1, p2) = (1.12e-07, -1.34e-03,  4.05), while the figure at the bottom shows the residuals in the approximation of the trend by the above polynomial. The green dashed line shows the linear fit for the data.}
\label{fa3}
\end{figure}

We performed a similar calibration for alpha-elements and found a temperature gradient with a difference of $\alpha$- abundances. Since Mg is the strongest tracer for $\alpha$ abundances \citep{2020ApJ...900..179K,2022AJ....164...60S}, we analyse the [Mg/Fe] from the HARPS and the GAIA sample. The Figure~\ref{fa3} shows the $\delta$[Mg/Fe] as a function of ${T_{eff}}$. We used the calibration procedure, similar to [Fe/H] and then applied the suitable corrections for the planet-hosting stars.

% \subsection{Monte Carlo analysis and K-S Test for stellar orbital parameters and Ages}
% The stellar orbital parameters and ages have uncertainties 

\end{appendix}
%\newpage

\bibliography{biblio}{}
\bibliographystyle{aasjournal}

\end{document}